\documentclass[aps,letter,preprint,amsfonts,amsmath,amssymb,prb]{revtex4}
\usepackage{amsmath,amsfonts,amssymb,bm}
\usepackage{array,multirow,dcolumn}
\usepackage{graphicx}
\usepackage{graphics}
\usepackage{color}
\usepackage{ulem}

\newcommand*{\vdf}[1]{\ensuremath{\text{d}\bm{#1}}}
\newcommand*{\vct}[1]{\ensuremath{\bm{#1}}}
\newcommand*{\norm}[1]{\ensuremath{\left|#1\right|}}
\newcommand*{\vnorm}[1]{\ensuremath{\left|\left|#1\right|\right|}}

\begin{document}

\title{A transferable \textit{ab-initio} based force field for aqueous ions}

\author{Sami Tazi}
\affiliation{UPMC Univ Paris 06, CNRS, ESPCI, UMR 7195 PECSA, F-75005 Paris, France}
\author{John J. Molina}
\affiliation{UPMC Univ Paris 06, CNRS, ESPCI, UMR 7195 PECSA, F-75005 Paris, France}
\author{Benjamin Rotenberg}
\affiliation{UPMC Univ Paris 06, CNRS, ESPCI, UMR 7195 PECSA, F-75005 Paris, France}
\author{Pierre Turq}
\affiliation{UPMC Univ Paris 06, CNRS, ESPCI, UMR 7195 PECSA, F-75005 Paris, France}
\author{Rodolphe Vuilleumier}
\affiliation{UPMC Univ Paris 06, Ecole Normale Sup\'erieure, D\'epartement de Chimie, F-75231 Paris, France}
\author{Mathieu Salanne}
\affiliation{UPMC Univ Paris 06, CNRS, ESPCI, UMR 7195 PECSA, F-75005 Paris, France}

\begin{abstract}
  We present a new polarizable force field for aqueous ions (Li$^+$, Na$^+$,
  K$^+$, Rb$^+$, Cs$^+$, Mg$^{2+}$, Ca$^{2+}$, Sr$^{2+}$ and Cl$^-$)
  derived from condensed phase \textit{ab-initio} calculations. 
  We use Maximally Localized Wannier Functions together with a generalized
  force and dipole-matching procedure to determine the whole set of parameters.
  Experimental data is then used only for validation purposes and 
  a good agreement is obtained for structural, dynamic and thermodynamic
  properties. The same procedure applied to crystalline phases allows to
  parametrize the interaction between cations and the chloride anion.
  Finally, we illustrate the good transferability of the force field 
  to other thermodynamic conditions by investigating concentrated solutions.
\end{abstract}

\maketitle

\section{Introduction}

The development of classical force fields for ions in aqueous solution
is essential to the description of specific effects, 
which are legion in biochemistry~\cite{heyda2010a,ingolfsson2011,domene2008},
atmospheric chemistry~\cite{jungwirth2006a} or environmental
science~\cite{rotenberg2009a}. 
The reliability of molecular simulations strongly depends on
the quality of the force field used to represent the interactions,
which must capture not only the effects of ionic size, but also 
the polarization of water by the ionic charge. The latter
multi-body effect becomes essential when dealing with 
multivalent ions~\cite{piquemal2006b},
in concentrated solutions~\cite{tribello2009a} 
and in interfacial environments~\cite{jungwirth2006a,chang2006a,wick2007a}. 

In recent years, a successful strategy to derive polarizable force fields
for solid and molten oxides from \textit{ab-initio} simulations has been
developed by Madden and
co-workers~\cite{aguado2003b,marrocchelli2010a,burbano2011a}. 
A full set of parameters was obtained
for the Ca-Mg-Al-Si-O (CMAS) system~\cite{jahn2007b}, which is the main
component of the Earth's crust and mantle. Cation-rich aluminosilicates,
including clays and zeolites, are
also the principal minerals on the Earth's surface, where they are in contact
with ionic solutions. Examples of situations where the interface between such
minerals and solution play an important role include the crystallization 
and dissolution of ionic crystals, such as calcium carbonate (in the context
of carbon dioxide sequestration~\cite{newell2008a}) or sodium sulfate (deterioration of
monuments~\cite{espinoza2010a}), or the sorption of radioactive contaminants 
(e.g. cesium or strontium) onto clays~\cite{glaus2007a,gimmi2011a}.
It is therefore of primary importance to
extend the CMAS force field in order to describe these minerals and their
interaction with water and ions. As a first step in this direction, 
we develop here a polarizable force field for ions in water which is
compatible with that developed for the CMAS system. 

As mentioned previously, and despite the success of
non-polarizable water force fields in the bulk
\cite{jorgensen1983a,berendsen1987a,abascal2007a,vega2011a}, transferability
to interfaces, especially charged ones, requires resorting to a polarizable model.
Many such models exist, which differ mainly in their treatment of the
polarizability. Drude or shell models assign a
charge on a spring to each polarizable
atom\cite{maaren2001,hansson2003,mackerell2003}. Other approaches allow 
for charge fluctuations \cite{sprik1991a,stuart1994} or assign point
dipoles to each polarizable species~\cite{caldwell1995a,berne1993,dang1997a,chialvo1996,sala2010a}. 
Only the latter model is compatible with the above-mentioned one for oxides.
Among the point polarizability models, we chose the one of Dang and Chang 
which was specifically developed
to describe the gas-liquid interface~\cite{dang1997a}. 
In addition, Masia \textit{et al.} have shown that it accurately reproduces 
the strong water polarization by divalent cations~\cite{masia2004a,masia2005a}.

Following the strategy of Madden and co-workers, which has proven able to
simultaneously reproduce structural, dynamic and thermodynamic properties
not only for the CMAS system, but also for many 
other ionic
materials~\cite{marrocchelli2010a,burbano2011a,heaton2006a,ohtori2009a},
we derive here the parameters of a force field for the aqueous ions: (Li$^+$,
Na$^+$, K$^+$, Rb$^+$, Cs$^+$, Mg$^{2+}$, Ca$^{2+}$, Sr$^{2+}$ and Cl$^-$.
The route from condensed phase Density Functional Theory (DFT) calculations, using 
Maximally Localized Wannier Functions (MLWFs)~\cite{marzari1997a,rotenberg2010a}
together with a generalized force and dipole-matching
procedure~\cite{madden2006a,salanne2012b}, renders experimental input
unnecessary, contrary to many force field parametrizations.

The paper is organized as follows: We first give a detailed
description of the force field and its parametrization which involves
DFT calculations on single ions in bulk water and on ionic crystals.
The second part is then devoted to the validation
of the model, against structural, dynamic and
thermodynamic properties of these systems. 
Finally, the transferability of the model is illustrated 
by the study of concentrated salt solutions.

\section{The force field and its parametrization}

\subsection{Model}

The total energy of the system is decomposed into four terms:
\begin{equation}
  V_\textrm{tot} = V_{\text{charge}}  + V_{\text{disp}} +
  V_{\text{rep}} + V_{\text{pol}}
\end{equation}
For the calculation of the direct Coulomb interaction between two
atoms $I$ and $J$,
\begin{equation}
  V_\textrm{charge} = \sum_{I,J>I} \frac{q^Iq^J}{r_{IJ}}
\end{equation}
formal charges (here $-1$, $+1$ or $+2$) are used.  The dispersion
potential includes the dipole-dipole and dipole-quadrupole terms
\begin{equation}
  \label{eq:disp}
  V_\textrm{disp} = - \sum_{I,J>I} \left[
    f_6^{IJ}(r_{IJ})\frac{C_6^{IJ}}{r_{IJ}^6}+
    f_8^{IJ}(r_{IJ})\frac{C_8^{IJ}}{r_{IJ}^8} \right]
\end{equation}
and the short-range corrections are described using the Tang-Toennies
functions $f_n^{IJ}$, which are of the form~\cite{tang1984a}:
\begin{equation}
  \label{eq:dampingdisp}
  f_n^{IJ} = 1-e^{-b_D^{IJ}r_{IJ}}\sum_{k=0}^{n} \frac{(b_D^{IJ}r_{IJ})^k}{k!}
\end{equation}
While the repulsion potential is modelled using a decaying exponential:
\begin{equation}
  \label{eq:rep}
  V_\textrm{rep} = \sum_{I,J>I} A^{IJ}e^{-B^{IJ}r_{IJ}}
\end{equation}
Finally, many-body electrostatic effects are described by the induced
dipoles $\mu^{I}$, which are treated as additional degrees of freedom
and obtained at each MD step by minimizing the polarization energy:
\begin{equation}
  V_\textrm{pol} = \sum_I\frac{1}{2\alpha^I}\norm{\mu^I}
  +\sum_{I,J}\left[\left(q^I\mu^J_{\alpha}g^{IJ}(r_{IJ})-q^J\mu^I_{\alpha}g^{JI}(r_{IJ})\right)T^{\alpha}_{IJ}
    -\mu^I_{\alpha}\mu^J_{\beta}T^{\alpha \beta}_{IJ}\right]
  \label{eq:pola}
\end{equation}
with $\alpha^I$ the ion polarizability and where the Einstein
summation convention is assumed.  A short-range correction to the
multipolar expansion of the Tang-Toennies type is used:
\begin{equation}
  \label{eq:dampingdipole}
  g^{IJ}(r_{IJ}) = 1-c^{IJ} e^{-b^{IJ}r_{IJ}}\sum_{k=0}^{4}
  \frac{(b^{IJ}r_{IJ})^k}{k!}
\end{equation}

This so-called Polarizable Ion Model (PIM) has proven extremely
successful for the description of oxides, chloride and fluoride-based
materials, both in the solid and liquid
states~\cite{marrocchelli2010a,burbano2011a,heaton2006a}. Water is
described by a model compatible with this form, developed by Dang and
Chang~\cite{dang1997a}. The only differences with the PIM are the
description of the repulsive and dispersion terms
$V_\textrm{rep}+V_\textrm{disp}$ for the water-water interactions,
represented by a Lennard-Jones potential, and the
absence of short-range damping of the charge-dipole interaction. The
Dang-Chang (DC) water is a rigid 4-site model, with an additional
virtual site M along the symmetry axis of the molecule, which bears
a negative partial charge, as well as the induced dipole, while the
Lennard-Jones interaction acts on the oxygen atom only. The parameters
of the DC model are summarized in Table~\ref{tab:dangchang}.

\begin{table}[ht]
\begin{center}
\begin{tabular}{|l|c|c|c|c|c|c|c|}
\hline
d$_\textrm{OH}$~\AA & d$_\textrm{OM}$~\AA & angle~($^{\circ}$) &
$\epsilon_\textrm{O}$
(kcal/mol) & $\sigma_\textrm{O}$~\AA & q$_\textrm{H}$ & $\alpha_\textrm{M}$~\AA\\\hline
\hline
0.9752 & 0.215 & 104.52 & 0.1825 & 3.2340 & 0.5190 & 1.444 \\\hline
\end{tabular}
\caption{\label{tab:dangchang} Parameters of the Dang-Chang water model.
}
\end{center}
\end{table}

The purpose of the present work is to derive all the parameters of the
PIM for water-ion and ion-ion interactions, thereby providing a force
field for the simulation of ions which is transferable from infinite
dilution to concentrated solutions, up to the ionic crystals, for
alkaline (Li$^+$, Na$^+$, K$^+$, Rb$^+$, Cs$^+$) and alkaline earth
(Mg$^{2+}$, Ca$^{2+}$, Sr$^{2+}$) cations and the chloride (Cl$^-$)
anion. Overall, this requires specifying 241 parameters. 
The procedure to determine all of them from \textit{ab-initio} calculations aims at
minimizing the risk of compensation of errors among the different
terms by 1) directly computing as many parameters as possible, 2)
adjusting the remaining ones on different quantities (dipoles
and forces) and 3) resorting to simplifying assumptions when
necessary.  We now describe these three aspects.

\subsection{Calculating parameters}
\label{sec:calculating}

\begin{figure}[ht]
  \begin{center}
  \includegraphics[scale=0.30]{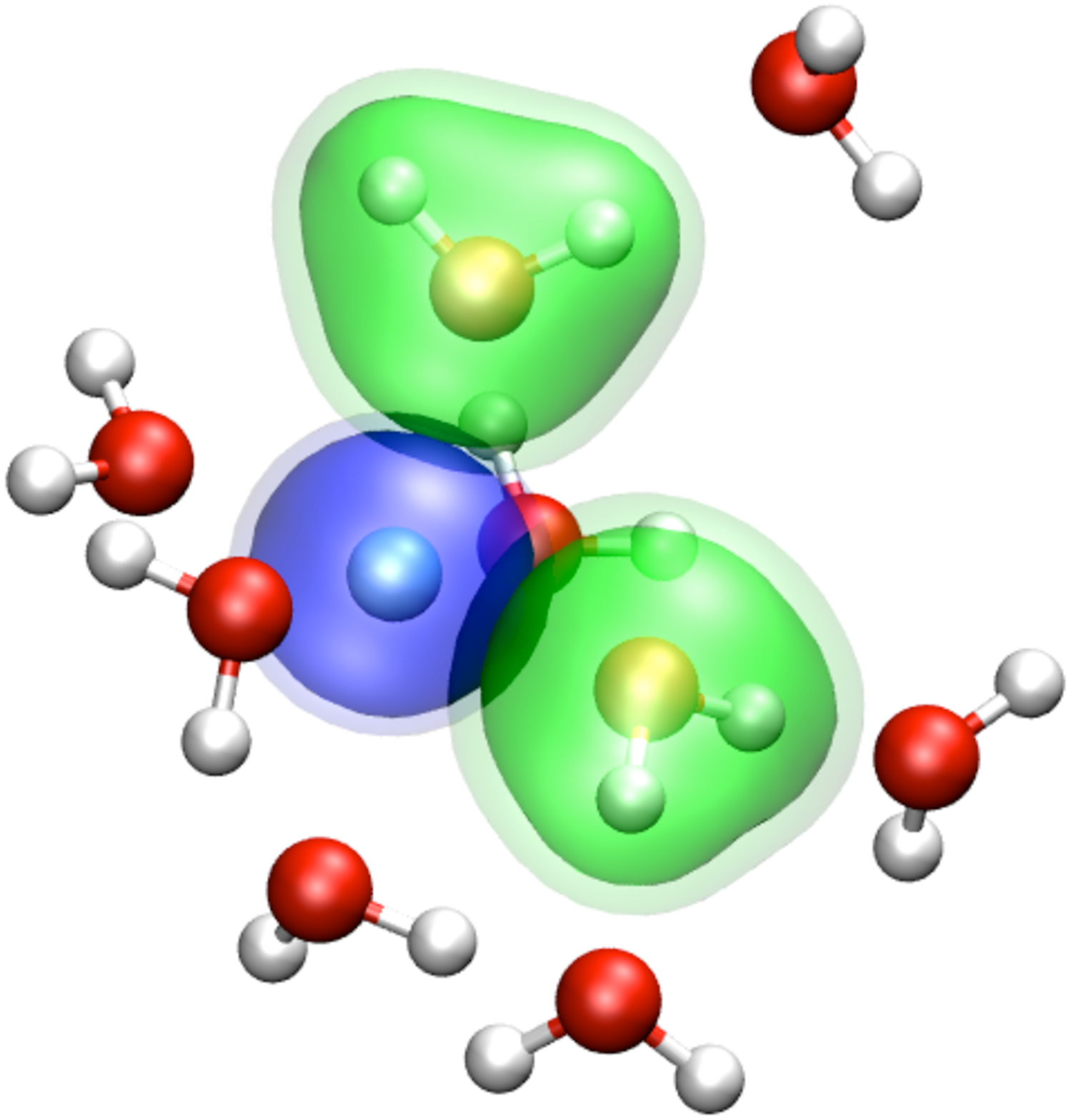}
  \caption{\label{fig:wannier} 
    Localized electronic density around a Ca$^{2+}$ cation and two water
    molecules in bulk water, reconstructed from their respective 
    Wannier orbitals. The isodensity surfaces include 90~\% and 95~\% of the 
    corresponding densities, respectively.}
  \end{center}
\end{figure}

First-principle calculations based on Density Functional Theory (DFT)
describe the electronic density 
using the Kohn-Sham orbitals, whose delocalized nature renders the
assignment of atomic or molecular properties difficult.  The concept
of the maximally localized Wannier function (MLWF) provides a
convenient framework to analyze atomic and molecular properties in the
condensed phase~\cite{martin2004a}. The Wannier functions are defined
through a unitary transformation of the Kohn-Sham eigenvectors.  MLWFs
are contructed by choosing the phase so that it minimizes the spread
of the Wannier function~\cite{martin2004a}.
It was shown recently that
MLWFs could be used to systematically derive both the polarizabilities
$\alpha^I$ and dispersion parameters $C_6^{IJ}$ and $C_8^{IJ}$ of a
PIM\cite{rotenberg2010a,salanne2012b}. Figure~\ref{fig:wannier} illustrates the electronic density 
around a Ca$^{2+}$ cation and two water molecules in bulk water, 
reconstructed from their respective Wannier orbitals.

\subsubsection{Polarizability}

In a closed shell system, each MLWF contributes two electrons, so that
the atomic or molecular dipole can be computed (in atomic units) as
\begin{equation}
\vct{\mu}^I=\sum_{i\in I}(Z_i\vct{R}_i-2\sum_{n\in i}\vct{r}^w_n)
\end{equation}
where the sums run over atoms $i$ belonging to fragment $I$ and over
MLWFs $n\in i$ whose center is localized in the vicinity of the
nuclear position $\vct{R}_i$, $Z_i$ is the charge of nucleus $i$ and
$r^w_n$ is the position of the center of the $n$-th MLWF. 
The polarizability may differ from that of the
same ion in the gas phase because of environmental effects.
It can be calculated by applying a small
electric field $\vct{\mathcal{E}}^{(\alpha)}$ along each Cartesian
direction $\alpha=x,y,z$ to the system, which induces dipole moments
$\left\{\delta\vct{\mu}^{I,(\alpha)}\right\}_{I\in[1,N]}$.  A
convenient way to distinguish the effect of the applied field from
that of the static fields caused by the permanent charge distributions
of the molecules, is to think of the former as an optical
field. $\delta\mu^I$ can then be seen as the net induced dipole
oscillating at the optical frequency. The total field
$\vct{f}^{I,(\alpha)}$ on each atom is
\begin{equation}
\label{eq:deltamu}
\vct{f}^{I,(\alpha)}=
\vct{\mathcal{E}}^{(\alpha)}+\sum_{J\ne I}\vct{T}^{IJ}\cdot
\delta\vct{\mu}^J({\vct{R}^N}) 
\end{equation}
where $\vct{T}^{IJ}$ is the dipole-dipole interaction tensor.  The
polarizability tensor of molecule $I$ can then be obtained by
inverting Eq.~(\ref{eq:deltamu})~:
\begin{equation}
  \vct{\alpha}^I({R^N})=(\vct{F}^I)^{-1}\cdot\vct{\Pi}^I
\end{equation}
with the second-rank three dimensional tensors defined as~: 
$F^I_{\alpha\beta}=f_\alpha^{I,(\beta)}$ 
and $\Pi^I_{\alpha\beta}=\delta\mu_\alpha^{I,(\beta)}$
More details about this approach can be found in
Ref.~\cite{molina2011a}.

\subsubsection{Dispersion: $C_6$ and $C_8$}

DFT calculations do not usually account for dispersion interactions,
because the former describe the electronic ground state while the
latter arise from correlated density fluctuations associated with
excited states.  The treatment of dispersion via non-local functionals
has only recently been introduced, albeit at a high computational
cost. Thus these interactions are generally added (if at all) as an \textit{a
posteriori} correction. Among the several methods that have been
proposed for computing this correction, the method of
Grimme~\cite{grimme2004a} and that of
Silvestrelli\cite{silvestrelli2008a} seem to be the most popular. In
this work we use the latter, which considers the dispersion
interaction between all pairs of MLWF as follows. The long-range
interaction between separated fragments of matter is calculated,
following Andersson et al.\cite{andersson1996a}, as
\begin{align}
E^{lr}_{xc}=\frac{6e}{4(4\pi)^{3/2}m^{1/2}}&\underset{V_1\,V_2}{\iint}
\vdf{r}_1\vdf{r}_2
\frac{\sqrt{\rho_1(\vct{r}_1)\rho_2(\vct{r}_2)}}{\sqrt{\rho_1(\vct{r}_1)}
  + \sqrt{\rho_2(\vct{r}_2)}}
\times\frac{1}{\vnorm{\vct{r}_1-\vct{r}_2}^6}
\label{eq:andersson} 
\end{align}
where  $\rho(r_i)$ is the charge density of
fragment $i$, $m$ the electron mass and $V_i$  the volume
occupied by fragment $i$.  For large separations $R$, this scales
as $E^{lr}=-C_6/R^6$, where the $C_6$ coefficient for the
interaction between two MLFWs $k$ and $l$ can be computed as~:
\begin{align}
C^{kl}_6=\frac{3}{32\pi^{3/2}}&\underset{\substack{r_1\le
    r_c\\r_2 \le
r_c'}}{\iint}\vdf{r}_1\vdf{r}_2\frac{w_k(\vct{r}_1)w_l(\vct{r}_2)}{w_k(\vct{r}_1)+w_l(\vct{r}_2)}
\label{eq:C6kl}
\end{align}
The cut-off radius $r_c= (1.475-0.866\ln{S})S$ is chosen to
correctly capture the limit of 
long-range perturbations in an electron
gas~\cite{silvestrelli2008a}.  The MLWFs, giving rise to densities
$\rho=w^2$, are assumed to be of the Slater form:
\begin{equation}
w_n(\vnorm{\vct{r}-\vct{r}_n}) = \frac{3^{3/4}}{\sqrt{\pi}S_n^{3/2}}
e^{-(\sqrt{3}/S_n)\vnorm{\vct{r}-\vct{r}_n}}
\end{equation}
characterized solely by their spread $S_n=\langle
w_n|r^2|w_n\rangle-\langle w_n|r|w_n\rangle^2$ and center $r_n$.

We have previously
shown that the dispersion interaction between two ensembles of charge
density fragments can be obtained from the averaged sum over pair
interactions of MLWFs~\cite{rotenberg2010a}. Assuming an isotropic
distribution of MLWF centers around the nuclei $I$ and $J$, at fixed
distances, leads (to second leading-order) to
$V_{disp}=-\sum_{n=6,8}C_n^{IJ}/r_{IJ}^n$, where the dispersion
coefficients are:
\begin{align}
C_6^{IJ}&=\sum_{k\in I,l\in J}C_6^{kl}\\
C_8^{IJ}&=\sum_{k\in I,l\in J}5(d^2_k+d^2_l)C_6^{kl}
\end{align}
where $d_{k,l}$ are the distances of the MLWF centers to their
respective nuclei and $C_6^{kl}$ is computed for each pair of MLWFs
according to Eq.~\ref{eq:C6kl}.  The determination of the parameter
$b_D$ in Eq.~\ref{eq:dampingdisp}, for the short-range damping of the
dispersion interaction, is detailed below.

\subsection{Dipole- and force-fitting}
\label{sec:matching}

Not all parameters of the force field can be derived systematically
from the electronic density. However, they can be determined
numerically so as to best reproduce the atomic properties calculted by
DFT: the total dipoles (permanent plus induced) of ions and molecules and the forces acting on
them.

\subsubsection{Damping of charge-dipole interaction}

The first step in our parametrization procedure is to
determine the parameters involved in Eq.~\ref{eq:dampingdipole} for
the short-range damping of the charge-dipole interaction. This is
achieved by numerically adjusting these parameters so as to minimize
the error on the dipoles calculated using the classical force field,
relative to the DFT ones on a number of representative configurations:
\begin{equation}
  \chi^2_{\mu} = \frac{1}{N_\textrm{conf}}  \frac{1}{N_\textrm{atom}}
  \sum_\textrm{conf}\ \sum_\textrm{atom} 
  \frac{ || \vct{\mu}^{classical} -
    \vct{\mu}^{DFT} ||^2 }  
  { || \vct{\mu}^{DFT} ||^2 }
\label{eq:chi2mu}
\end{equation}
Together with the polarizabilities, these parameters complete the
description of the polarization potential $V_\textrm{pol}$.

\subsubsection{Repulsion}

The parameters of the repulsive potential $V_\textrm{rep}$ in Eq.~\ref{eq:rep} 
can then be obtained by a similar procedure as the one used for the dipoles,
if the functional used for the DFT calculation does not include
dispersion interactions (e.g. PBE or BLYP)~\cite{jones1989a,kristyan1994a}:
\begin{equation}
  \chi^2_{F} = \frac{1}{N_\textrm{conf}}  \frac{1}{N_\textrm{atom}}
  \sum_\textrm{conf}\ \sum_\textrm{atom} 
  \frac{ || \vct{F}^{classical} - \vct{F}^{DFT} ||^2 } { || \vct{F}^{DFT} ||^2 }
\label{eq:chi2}
\end{equation}
By ajusting the parameters for the damping of the charge-dipole
interaction and for the repulsion on different physical quantities
(dipoles and forces, respectively), we limit the risk of having a
compensation of errors between the different terms of the potential.

\subsection{Further considerations}

The water-ion interactions are parametrized by applying the procedure
described above on configurations of a system containing a
single ion in bulk water. For the ion-ion interactions, we
use configurations of the experimentally stable crystal phase under
normal conditions: NaCl structure for Li$^+$, Na$^+$, K$^+$ and
Rb$^+$, CsCl structure for Cs$^+$, MgCl$_2$ structure for Mg$^{2+}$,
and CaCl$_2$ structure for Ca$^{2+}$ and Sr$^{2+}$.  The Cl-Cl
interactions must be the same among the different crystals in order to
ensure the consistency and transferability of our potentials. The
parameters for the Cl-Cl repulsion are obtained for LiCl, in which
they are the most prominent, and the corresponding values are then
used for all crystals. The $C_6$ and $C_8$ parameters for the Cl-Cl
dispersion interaction are obtained by averaging the values for the
different crystals.

For the cation-anion repulsion (see Eq.~\ref{eq:rep}), the
force-fitting procedure results in $B$ parameters that are very close
to each other among the alkaline ions on the one hand, and among the
alkaline earth ions on the other hand. For the sake of simplicity, we
use only one value for this parameter for each ion series. The $A$
parameters for the cation-anion repulsion are then readjusted to
minimize Eq.~\ref{eq:chi2} while keeping the $B$ value fixed.  The
final values for $A$ and the corresponding $\chi^2_F$ were practically
unchanged by this constraint, thus confirming the relevance of this
choice.

In order to further decrease the number of free parameter, 
the range of the short-range damping
used for the cation-anion dispersion $b_D^{IJ}$ (see
Eqs.~\ref{eq:disp} and~\ref{eq:dampingdisp}) is taken in most cases
equal to that of the short-range repulsion $B^{IJ}$.  This assumption
is not new~\cite{stone-book}, and it is justified by the notion that
the long-range scaling of dispersion breaks down as the electronic
fragments start overlapping, when the short-range repulsion comes
into play.  The damping of the Cl-Cl dispersion is adjusted
numerically so as to reproduce simultaneously the experimental
density of all crystals.  For the largest cations, Cs$^+$ and
Sr$^{2+}$, a value slightly smaller than $B^\textrm{cation-Cl}$ was needed to
reproduce the experimental densities.  Compared to the usual procedure
of parametrizing a PIM from \textit{ab-initio}
simulations~\cite{heaton2006a,marrocchelli2010a}, the systematic
determination of the $C_6$ and $C_8$ coefficients and the assumption
that $b_D=B$ dramatically reduce the number of parameters that need
to be adjusted in order to reproduce the whole set of experimental
densities. The damping parameter of the monovalent cation-water dispersion
interaction was chosen equal to that for the corresponding monovalent
cation-Cl$^-$ dispersion interaction
$b_D^\textrm{ion-O}=b_D^\textrm{ion-Cl}$, since the water molecules 
and the Cl$^-$ ions have approximately the same size.
As far as the divalent cations are concerned,
the attractive force arising at short distances from dispersion
is negligible compared to the charge-charge and charge-dipole
interactions.  We can thus omit damping this interaction without
any loss of accuracy.
Similarly, the dispersion interaction between Cl$^-$ and water oxygen atom 
is not damped.

Overall, these considerations reduce the number of parameters for the
interaction of all ions with water and of cations with chloride from 241 to 187, 
after the neglect of some terms for the reasons explained above, and to 170 by
further assuming that the ranges of some interactions are equal.
Out these 170, only 82 are adjusted numerically from the dipole- and force-matching
procedures of section~\ref{sec:matching}, while the rest are computed as expained
in section~\ref{sec:calculating}.

\subsection{Simulation details}

The parametrization of the force field from \textit{ab-initio} simulations is
achieved using representative configurations of the aqueous ions and
the ionic crystals.  For each ionic species, $\sim100$ configurations of
a system containing a single ion and 32 water molecules are generated
using the force-field of Dang \textit{et al.}~\cite{dang1992a,chang1997a,dang2002a} 
for the monovalent ions, and
that of Yu \textit{et al.}~\cite{yu2010a} for the divalent ions.  DFT
calculations were then performed on these configurations with the BLYP
functional~\cite{becke1988a,lee1988a} (exept for the Rb$^+$, for which the PBE
functional~\cite{perdew1996a} was used). The
Troullier-Martins\cite{troullier1991a} (Cl$^-$,Cs$^+$ and K$^+$) and
Goedecker-Teter-Hutter\cite{goedecker1996a,hartwigsen1998a,krack2005a}
(Na$^+$, Rb$^+$, Mg$^{2+}$, Ca$^{2+}$ and Sr$^{2+}$) pseudopotentials
were used, with a plane-wave basis set and an energy cutoff of at
least 70~Ry.  Similarly, 
configurations of crystals containing between 16 and 108 MCl or MCl$_2$ units,
are used to perform the DFT calculations,
with the same functionals and pseudopotentials as for the ions in
water. In each case, after determining the electronic density, the
forces acting on each atom are computed and the dipoles are
calculated from the MLWFs as described above. The $C_6$ and $C_8$
dispersion parameters are computed from the spreads and distances to
the center of the MLWFs, which result from the localization procedure.
The polarizabilities are calculated as explained above, by applying an
external field using the Berry phase
representation~\cite{molina2011a}.  All \textit{ab-initio} calculations were
performed using the CPMD simulation package~\cite{cpmd} (exept for
those involving Ca$^{2+}$, performed with CP2K simulation
package\cite{cp2k}), while classical forces and dipoles are computed
on the same configurations with FIST, the classical MD module of the
CP2K simulation package~\cite{cp2k}.  The numerical minimization of
Eqs.~(\ref{eq:chi2mu}) and~(\ref{eq:chi2}) is performed using the
Minuit library~\cite{james1975a}.

\subsection{Parametrization: Results}

The computed polarizabilities for all the ions are summarized in table
\ref{tab:pola}.  As expected, the polarizability increases when going
down along columns of the periodic table (alkaline and alkaline earths
series), while a decrease is observed when going from left to right
along rows (Na$^+$ to Mg$^{2+}$, K$^+$ to Ca$^{2+}$ and Rb$^+$ to
Sr$^{2+}$). For cations, the condensed phase polarizability is
comparable to that in the gas phase, except for Cs$^+$.  For the
chloride anion, however, the confinement of electrons by the
surrounding water molecules results in a significant decrease of the
polarizability (approximately 35\%).  A more detailed discussion has
been given in Ref.~\cite{molina2011a}.


\begin{table}[ht]
\begin{center}
\begin{tabular}{|l|c|c|c|}
  \hline
  Ion & $\alpha$ (\AA$^{3}$) & $\sqrt{\langle\vct{\mu}^2\rangle}$ (Debye)\\\hline
  \hline
  Li$^+$    & 0.03 &  0.002 \\\hline
  Na$^+$    & 0.18 &  0.014 \\\hline
  K$^+$     & 0.81 &  0.062 \\\hline
  Rb$^+$    & 1.32 &  0.097 \\\hline
  Cs$^+$    & 2.02 &  0.153 \\\hline\hline
  Mg$^{2+}$ & 0.08 &  0.010 \\\hline
  Ca$^{2+}$ & 0.44 &  0.026 \\\hline
  Sr$^{2+}$ & 0.81 &  0.071 \\\hline
  \hline
  Cl$^-$    & 3.50 &  0.415 \\\hline
\end{tabular}
\caption{\label{tab:pola} Polarizability $\alpha$ and magnitude of the induced
dipole of each ion $\sqrt{\langle\vct{\mu}^2\rangle}$. The latter
is 1.18~D for water. }
\end{center}
\end{table}

Interestingly, as indicated in Table~\ref{tab:pola}, the (induced)
dipole moment of cations is always very small compared to that of the
chloride anion and water.  This can be explained by the combination of
two factors.  First, most cations have a small (K$^+$, Sr$^{2+}$) or
very small (Li$^+$, Na$^+$, Mg$^{2+}$, Ca$^{2+}$) polarizability.
Second, all cations have a highly symmetric hydration sphere, which
results in very weak local electric fields to polarize them. Because
the induced dipoles are very small, they are not easily reproduced by
the classical force field (typical errors are of the order of 100\%),
but they do not contribute significantly to the polarization energy
$V_\textrm{pol}$, which is dominated by the interaction of the ionic charge
with the dipole of water, and hence to the forces.  For the sake of
simplicity, we thus decided to neglect the polarizability of all
cations and not include any additional degrees of freedom to describe
their induced dipoles.

\begin{table}[ht]
\begin{center}
\begin{tabular}{|l|c|c|c|c|c|c|c|c|c|}
  \hline
  System & A$^\textrm{ion-O}$ (Ha) & B$^\textrm{ion-O}$ (\AA$^{-1}$) &
  C$_6^\textrm{ion-O}$ (Ha.\AA$^{6}$) & C$_8^\textrm{ion-O}$ (Ha.\AA$^{8}$) &
  b$_D^{IJ}$ (\AA$^{-1}$)& b$^\textrm{ion-M}$ (\AA$^{-1}$)  &
  c$^\textrm{ion-M}$   \\ \hline \hline
  Li$^{+}$-water & 24.75 & 4.094 & 1.103$\times$10$^{-2}$ & 1.037$\times$10$^{-2}$ & 3.000 & 4.011 & 2.950 \\\hline
  Na$^{+}$-water & 711.1 & 5.061 & 1.335$\times$10$^{-1}$ & 1.572$\times$10$^{-1}$ & 3.000 & 1.562 & 6.839$\times$10$^{-1}$ \\\hline
  K$^{+}$-water & 125.7 & 3.735 & 7.530$\times$10$^{-1}$ & 1.206 & 3.000 & 1.315 & 4.623$\times$10$^{-1}$ \\\hline
  Rb$^{+}$-water & 157.8 & 3.656 & 1.225 & 2.267 & 3.000 & 1.248 & 4.765$\times$10$^{-1}$ \\\hline
  Cs$^{+}$-water & 269.4 & 3.635 & 2.040 & 4.644 & 1.800 & 2.524 & 2.948 \\\hline\hline 
  Mg$^{2+}$-water & 65.67 & 3.963 & 6.408$\times$10$^{-2}$ & 7.23$\times$10$^{-3}$ & - & 3.963 & 2.820 \\\hline
  Ca$^{2+}$-water & 57.94 & 3.327 & 5.055$\times$10$^{-1}$ & 7.502$\times$10$^{-1}$ & - & 3.327 & 3.000 \\\hline
  Sr$^{2+}$-water & 41.55 & 2.991 & 9.159$\times$10$^{-1}$ & 1.576 & - & 2.991 & 2.041 \\\hline
\end{tabular}
\caption{\label{tab:ion-water} Parameters for the cation-water
  interactions.  As for water-water interactions, respulsion and
  dispersion involve the oxygen atom, while electrostatic interactions
  involve the additional M site.  The damping parameter $b_D$ for the
  dispersion interaction for the monovalent ions is chosen equal to
  that of the corresponding cation-chloride interaction (see text and
  table~\ref{tab:ionion}). 
  The electrostatic damping is between the water dipole and cation charge.
  }
\end{center}
\end{table}

\begin{table}[ht]
\begin{center}
\begin{tabular}{|l|c|c|c|c|c|c|c|c|}
  \hline
  System & A$^{\text{ion-O}}$ (Ha) & B$^\textrm{ion-O}$ (\AA$^{-1}$) 
  & C$_6^\textrm{ion-O}$ (Ha.\AA$^{6}$)  & C$_8^\textrm{ion-O}$ (Ha.\AA$^{8}$) &
  b$^\textrm{ion-H}$  (\AA$^{-1}$) & c$^\textrm{ion-H}$ 
  & b$^\textrm{ion-M}$ (\AA$^{-1}$)  & c$^\textrm{ion-M}$ 
  \\ \hline \hline 
  Cl-water & 499.63 & 3.560 & 2.039 & 4.296 & 4.794 & 1.093 & 2.444 &
  -1.901\\\hline
\end{tabular}
\caption{\label{tab:paramcl} Parameters for the chloride-water
  interactions. The dipole damping is between the Cl$^-$ and the water charges. For the reasons already explained, there is no damping of the dispersion.
    }
\end{center}
\end{table}

\begin{table}[ht!]
\begin{center}
\begin{tabular}{|l|c|c|c|c|c|c|c|c|}
  \hline
  System & Ion pair IJ& A$^{IJ}$ (Ha) & B$^{IJ}$ (\AA$^{-1}$)  & C$_6^{IJ}$
  (Ha.\AA$^{6}$)  &
  C$_8^{IJ}$ (Ha.\AA$^{8}$)  & b$_D^{IJ}$  (\AA$^{-1}$) & b$^{IJ}$  (\AA$^{-1}$)
  & c$^{IJ}$ \\ \hline \hline
  LiCl & Li$^+$-Li$^+$ & 481.9 & 6.958 & 2.727$\times$10$^{-4}$ &
  5.570$\times$10$^{-10}$ & 6.958 & - & -\\ 
  & Li$^+$-Cl$^-$ & 15.56 & 3.000 & 2.369$\times$10$^{-2}$ & 2.511$\times$10$^{-2}$ & 3.000 & 3.128 & 1.433 \\
  & Cl$^-$-Cl$^-$ & 698.4 & 3.777 & 5.951 & 12.85 & 1.650 & - & -\\\hline
  NaCl & Na$^+$-Na$^+$ & 1.701$\times$10$^{-2}$ & 4.965 & 2.914$\times$10$^{-2}$ & 1.394$\times$10$^{-2}$ & 4.965 & - & -\\ 
  & Na$^+$-Cl$^-$ & 44.43 & 3.000 & 2.971$\times$10$^{-1}$ & 3.785$\times$10$^{-1}$ & 3.000 & 2.775 & 2.040\\ 
  & Cl$^-$-Cl$^-$ & 698.4 & 3.777 & 5.951 & 12.85 & 1.650 & - & -\\\hline
  KCl  & K$^+$-K$^+$ & 174.9 & 5.000 & 7.172$\times$10$^{-1}$ & 9.260$\times$10$^{-1}$ & 5.000 & - & -\\
  & K$^+$-Cl$^-$ & 82.92 & 3.000 & 1.973 & 3.347 & 3.000 & 1.282 & 9.059$\times$10$^{-1}$\\ 
  & Cl$^-$-Cl$^-$ & 698.4 & 3.777 & 5.951 & 12.85 & 1.650 & - & -\\\hline
  RbCl & Rb$^+$-Rb$^+$ & 1.235$\times$10$^{-2}$ & 3.485 & 2.235 & 3.908 & 3.485 & - & -\\ 
  & Rb$^+$-Cl$^-$ & 108.0 & 3.000 & 3.755 & 7.223 & 3.000 & 1.460 & 9.825$\times$10$^{-1}$\\ 
  & Cl$^-$-Cl$^-$ & 698.4 & 3.777 & 5.951 & 12.85 & 1.650 & - & -\\\hline
  CsCl & Cs$^+$-Cs$^+$ & 353.0 & 3.782 & 7.325 & 18.64 & 3.782 & - & -\\ 
  & Cs$^+$-Cl$^-$ & 150.1 & 3.000 & 7.339 & 16.96 & 1.800 & 1.541 & 4.665$\times$10$^{-1}$\\ 
  & Cl$^-$-Cl$^-$ & 698.4 & 3.777 & 5.951 & 12.85 & 1.650 & - & -\\
  \hline\hline
  MgCl$_2$ & Mg$^+$-Mg$^{2+}$ & 2.231$\times$10$^{-1}$ & 4.995 & 
  1.095$\times$10$^{-2}$ & 4.066$\times$10$^{-3}$ & 4.995 & - & -\\ 
  & Mg$^{2+}$-Cl$^-$ & 85.84 & 3.400 & 1.471$\times$10$^{-1}$ & 2.102$\times$10$^{-1}$ & 3.400 & 2.886 & 2.113\\ 
  & Cl$^-$-Cl$^-$ & 698.4 & 3.777 & 5.951 & 12.85 & 1.650 & - & -\\\hline
  CaCl$_2$ & Ca$^{2+}$-Ca$^{2+}$ & 1.289$\times$10$^{-1}$ & 3.941 & 3.274$\times$10$^{-1}$ & 3.456$\times$10$^{-1}$ & 3.941 & - & -\\ 
  & Ca$^{2+}$-Cl$^-$ & 236.3 & 3.400 & 1.168 & 1.883 & 3.400 & 2.052 & 1.268\\ 
  & Cl$^-$-Cl$^-$ & 698.4 & 3.777 & 5.951 & 12.85 & 1.650 & - & -\\\hline
  SrCl$_2$ & Sr$^{2+}$-Sr$^{2+}$ & 5.513 & 4.735 & 1.259 & 1.867 & 4.735 & - & -\\ 
  & Sr$^{2+}$-Cl$^-$ & 269.4 & 3.400 & 2.697 & 5.066 & 2.400 & 3.103 & 2.939\\ 
  & Cl$^-$-Cl$^-$ & 698.4 & 3.777 & 5.951 & 12.85 & 1.650 & - & -\\\hline
\end{tabular}
\caption{\label{tab:ionion} Parameters for the ion-ion interactions.}
\end{center}
\end{table}

\begin{figure}[ht]
  \begin{center}
  \includegraphics[scale=0.30]{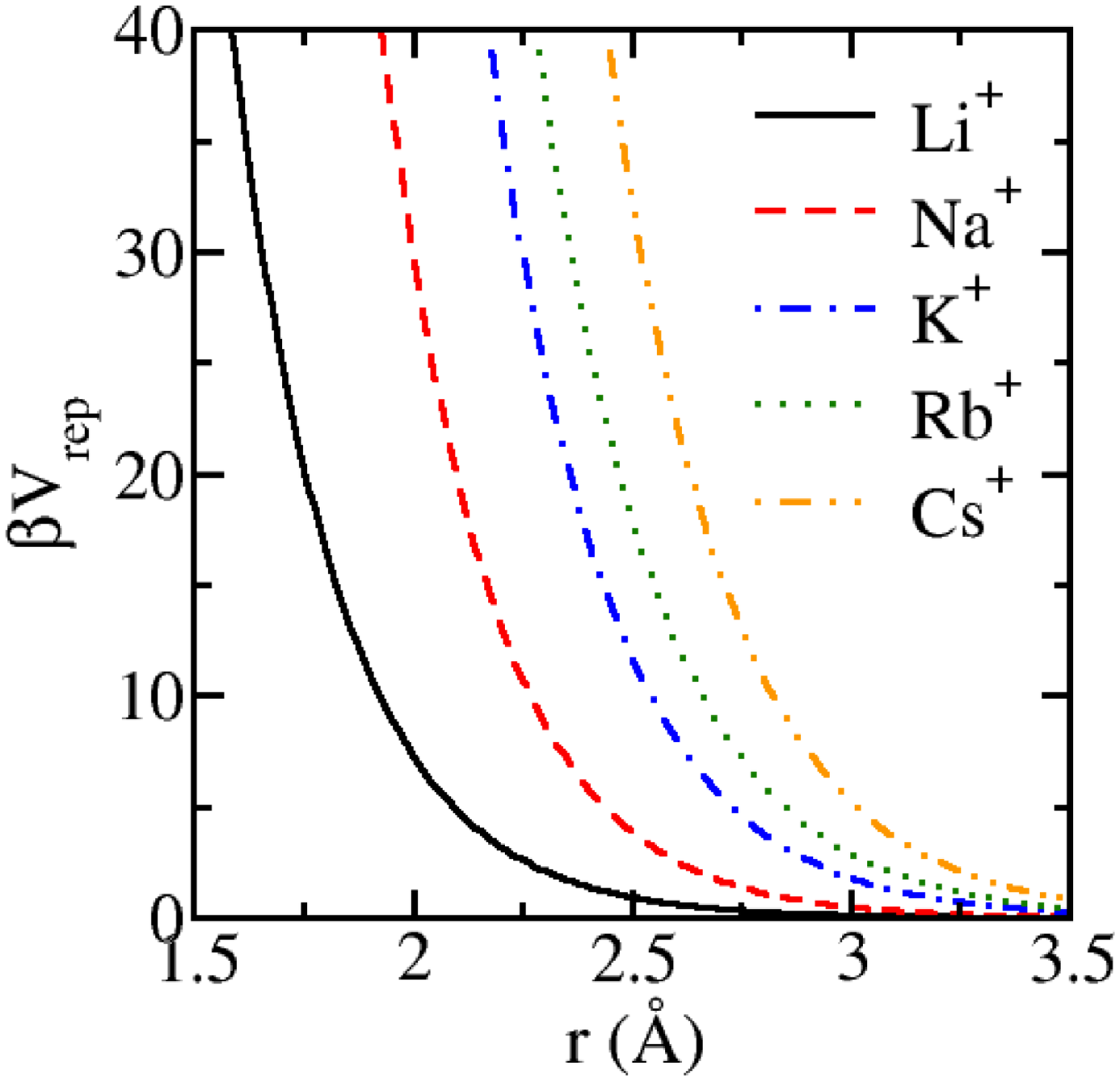}
  \includegraphics[scale=0.30]{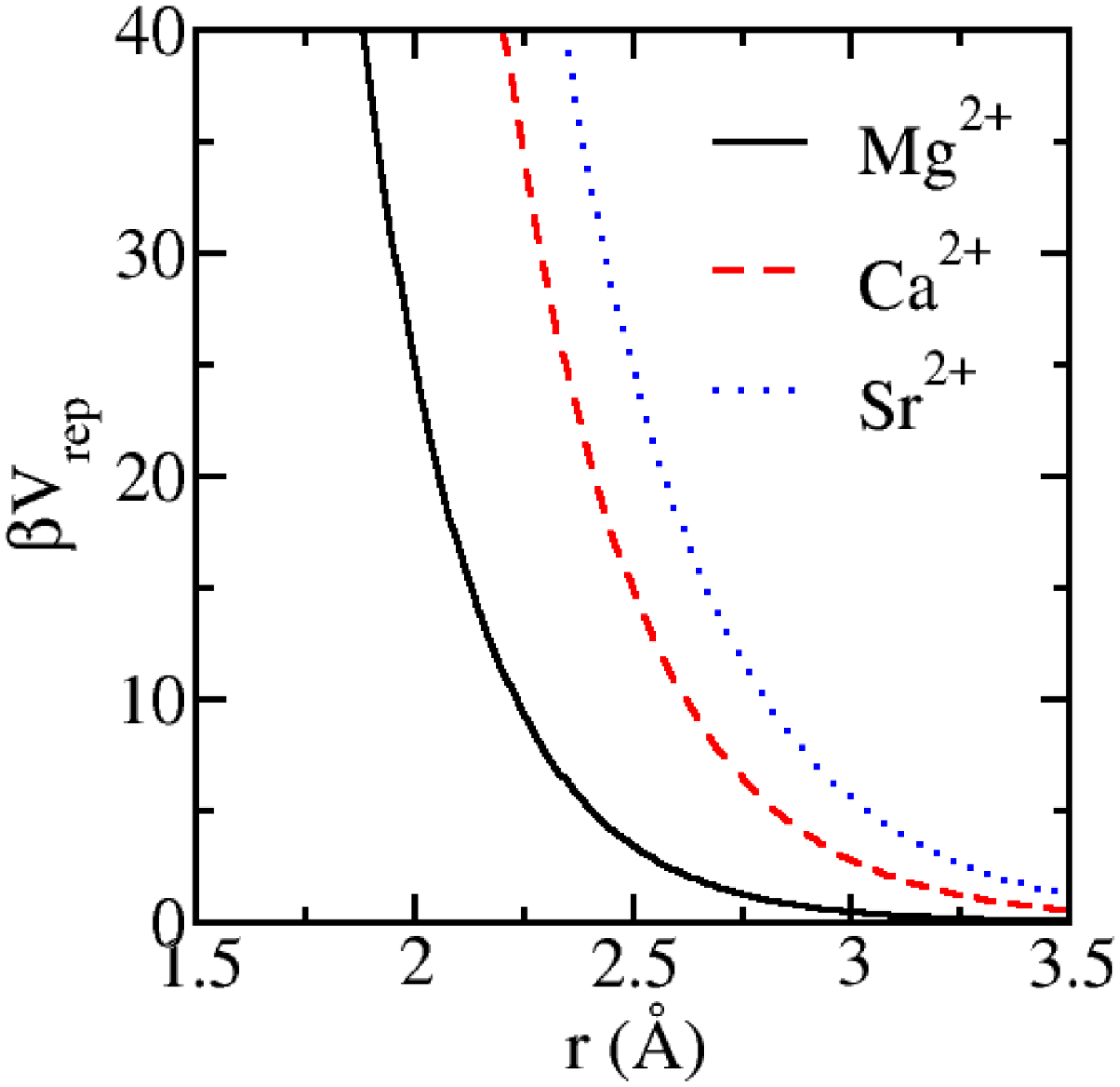}
  \caption{\label{fig:vrep} Repulsion potential between water and
    alkaline cations (a) and alkaline earth cations (b), in units of
    the thermal energy $\beta^{-1}=k_BT$.}
    \end{center}
\end{figure}

The parameters for the cation-water interaction are summarized in
Table~\ref{tab:ion-water}, and those for the chloride-water
interaction are given in Table~\ref{tab:paramcl}. Finally, all
parameters for the ion-ion interactions are given in
Table~\ref{tab:ionion}.  The resulting repulsion potentials
$V_\textrm{rep}$ between water and the various cations, plotted in
Fig.~\ref{fig:vrep}, nicely reflect the expected increase in ionic
size along the alkaline and alkaline earth series. Furthermore, a comparable
repulsion is observed for isoelectronic species such as Na$^+$ and
Mg$^{2+}$, K$^+$ and Ca$^{2+}$, and Rb$^+$ and Sr$^{2+}$.  In line
with the polarizabilities, the $C_6$ and $C_8$ dispersion coefficients
for the ion-water interaction increase along the alkaline and alkaline
earths series, while a decrease is observed from left to right along
rows of the periodic table. The same trends hold for the repulsion and
dispersion interactions between the cations and the chloride anion.

\begin{figure}[ht]
\begin{center}
\includegraphics[scale=0.40]{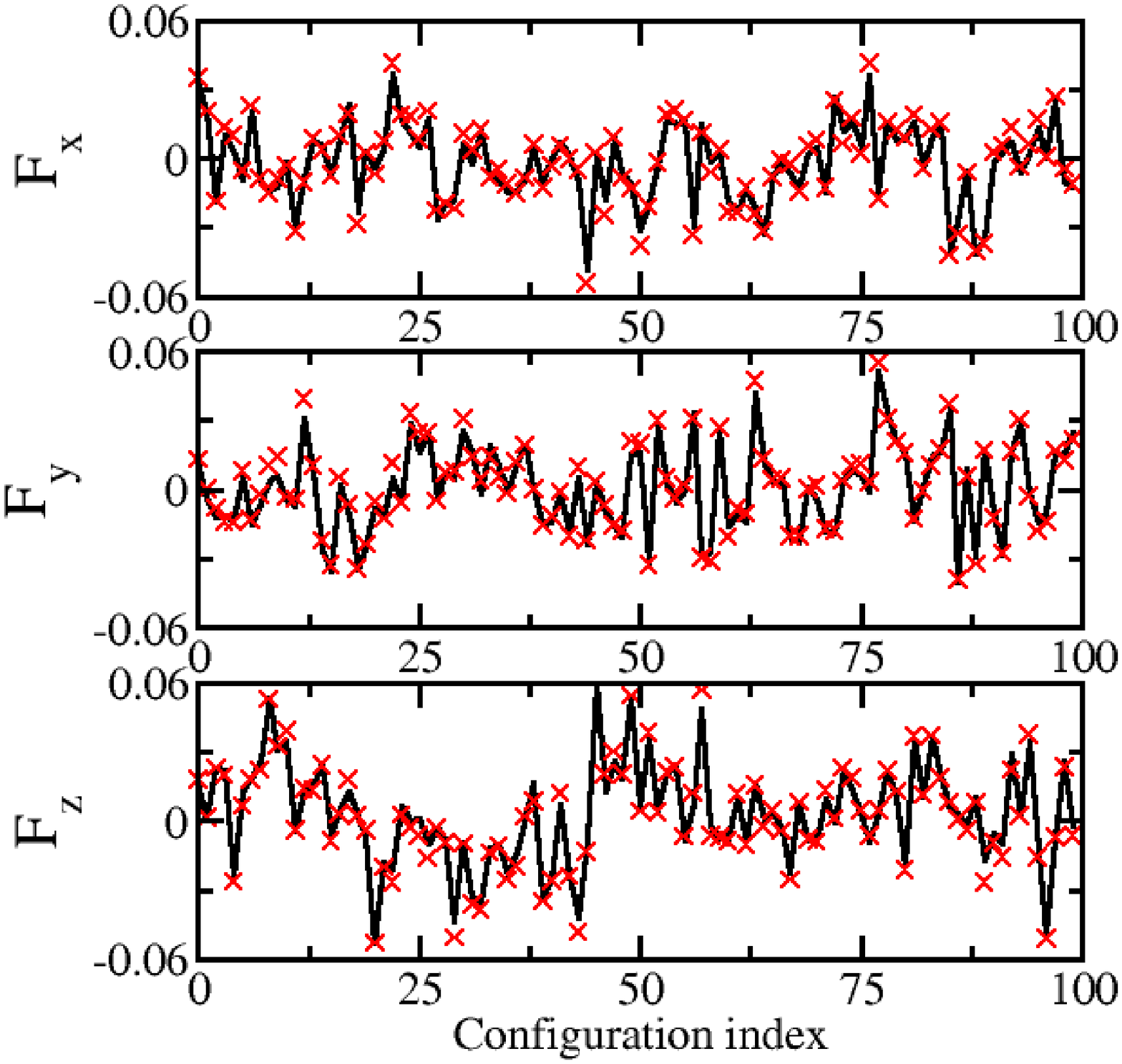}
\caption{\label{fig:DFTvsClassic} Force (in atomic units) acting on the Ca$^{2+}$ ion.
  The prediction of the classical force field (lines) for the 3
  components $F_x$, $F_y$ and $F_z$ are compared to the DFT result
  ($\times$), for 100 configurations composed of 32 water molecules
  and 1 Ca$^{2+}$. }
\end{center}
\end{figure}

We now examine the performance of the force field in terms of
reproducing the \textit{ab-initio} dipoles and forces.
Figure~\ref{fig:DFTvsClassic} illustrates the comparison between the
forces on the ion calculated with the classical force field (without
dispersion) and those obtained from the DFT calculations, for the
Ca$^{2+}$ cation. From 
table~\ref{tab:chi2ions}, the relative error of the force,  $\sqrt{\chi^2_F}$, on the
Ca$^{2+}$ ion, with respect to the DFT result, is approximately 23\% .
This can be considered as a good match, especially when comparing to
the corresponding results obtained by using the Dang
potential~\cite{dang2006a} (with the same water model), which results
in a relative error of approximately $320\%$. Similarly, our results
for Sr$^{2+}$ show a relative error of 36\%, compared to 501\% with
the force field from the literature~\cite{dang2003a}, 49\% vs. 66\%
for Na$^+$, 48\% vs. 131\% for Cs$^+$ and 53\% vs. 104\% for Cl$^-$.
Overall, the forces on all ions are well reproduced by the
present force field. The largest contributions to the
relative error (see Eq.~\ref{eq:chi2}) correspond to the smaller forces.

\begin{table}[ht]
\begin{center}
\begin{tabular}{|l|c|c|c|}
  \hline
  Ion & $\chi^{2}_\textrm{F-ion}$ & $\chi^{2}_\textrm{$\mu-H_2O$}$ &
  $\chi^2_\textrm{$\mu$-ion}$ \\ 
  \hline \hline
  Li$^{+}$ & 1.56$\times$10$^{-1}$ & 1.76$\times$10$^{-3}$ & - \\\hline
  Na$^{+}$ & 2.36$\times$10$^{-1}$ & 7.81$\times$10$^{-3}$ & - \\\hline
  K$^{+}$  & 1.11$\times$10$^{-1}$ & 2.79$\times$10$^{-3}$ & - \\\hline
  Rb$^{+}$ & 1.02$\times$10$^{-1}$ & 2.35$\times$10$^{-3}$ & - \\\hline
  Cs$^{+}$ & 2.28$\times$10$^{-1}$ & 3.08$\times$10$^{-3}$ & - \\\hline\hline 
  Mg$^{2+}$ & 9.97$\times$10$^{-2}$ & 1.78$\times$10$^{-2}$ & - \\\hline
  Ca$^{2+}$ & 5.35$\times$10$^{-2}$  & 9.05$\times$10$^{-3}$ & - \\\hline
  Sr$^{2+}$ & 1.27$\times$10$^{-1}$ & 2.73$\times$10$^{-3}$ & - \\\hline
  \hline 
  Cl$^-$ & 2.84$\times$10$^{-1}$  & 3.40$\times$10$^{-3}$  & 2.05$\times$10$^{-1}$ \\
  \hline 
\end{tabular}
\caption{\label{tab:chi2ions} $\chi^2$ for the forces on the ions and the
dipoles of water and the ions.}
\end{center}
\end{table}

Table~\ref{tab:chi2cryst} reports the $\chi^2$ values obtained on the
crystals for the forces on both cations and anions, as well as for the
dipole of the anions.  Comparison with Table~\ref{tab:chi2ions}
indicates that a similar accuracy is obtained for both the crystals
and the ions in solution, suggesting that the force field should
perform well under both conditions. This result is also encouraging
from the point of view of the transferability and the possible
prediction of the solubility of these crystals. Comparison between
tables~\ref{tab:chi2cryst} and~\ref{tab:chi2crystDC} illustrates 
the better performance of the present model compared to those of Dang
and coworkers~\cite{dang1992a,chang1997a,dang2002a}.

\begin{table}[ht]
  \begin{center}
    \begin{tabular}{|l|c|c|c|}
      \hline
      Crystal & $\chi^2_\textrm{F-M$^{x+}$}$ & $\chi^2_\textrm{F-Cl$^-$}$ &
      $\chi^2_\textrm{$\mu$-Cl$^-$}$\\ \hline \hline
      LiCl     & 1.13$\times$10$^{-1}$ & 1.74$\times$10$^{-2}$ & 1.91$\times$10$^{-1}$ \\\hline
      NaCl     & 2.52$\times$10$^{-2}$ & 1.12$\times$10$^{-2}$ & 1.87$\times$10$^{-1}$ \\\hline
      KCl      & 7.88$\times$10$^{-2}$ & 5.61$\times$10$^{-2}$ & 7.28$\times$10$^{-1}$ \\\hline
      RbCl     & 4.77$\times$10$^{-2}$ & 6.46$\times$10$^{-2}$ & 6.66$\times$10$^{-1}$ \\\hline
      CsCl     & 2.23$\times$10$^{-2}$ & 1.15$\times$10$^{-1}$ & 4.30$\times$10$^{-1}$ \\\hline\hline
      MgCl$_2$ & 2.62$\times$10$^{-1}$ & 8.45$\times$10$^{-2}$ & 1.90$\times$10$^{-2}$ \\\hline
      CaCl$_2$ & 5.58$\times$10$^{-2}$ & 8.46$\times$10$^{-3}$ & 2.27$\times$10$^{-1}$ \\\hline
      SrCl$_2$ & 3.61$\times$10$^{-2}$ & 3.95$\times$10$^{-2}$ & 5.69$\times$10$^{-2}$ \\\hline
    \end{tabular}
    \caption{\label{tab:chi2cryst} $\chi^2$ in crystals, for the
      forces on the cations and anions, and the dipoles of anions.}
  \end{center}
\end{table}

\begin{table}[ht]
  \begin{center}
    \begin{tabular}{|l|c|c|}
      \hline
      Crystal & $\chi^2_\textrm{F-M$^{x+}$}$ & $\chi^2_\textrm{F-Cl$^-$}$ \\\hline
      LiCl     & 3.51 & 28.0 \\\hline
      NaCl     & 4.46$\times$10$^{-1}$ & 3.50 \\\hline
      KCl      & 4.38 & 4.10 \\\hline
      CsCl     & 2.24 & 2.56 \\\hline\hline
      CaCl$_2$ & 1.03 & 9.85$\times$10$^{-1}$ \\\hline
      SrCl$_2$ & 5.47 & 8.28 \\\hline
\end{tabular}
\caption{\label{tab:chi2crystDC} $\chi^2$ in crystals, for the
  forces on the cations and anions, with the polarizable Dang-Chang
  models.}
\end{center}
\end{table}

Neglecting the polarizability of cations does not prevent us from
obtaining a good description of the forces acting on them, as can be
seen in table~\ref{tab:chi2ions}.  These forces are even better
described than those on the chloride ion, whose polarizability is
explicitly taken into account.  Nevertheless, for the reasons
mentioned in the introduction, it is essential to correctly reproduce
the polarization of water molecules around ions.
Table~\ref{tab:chi2ions} also indicates the relative error on the
dipole of water molecules in the first solvation shell of the
ions. The combination of the Dang-Chang water model with the present
model for the ion-water interactions provides a very good description
of the polarization of water, with relative errors between 5 and 10\%
for all ions except Mg$^{2+}$ (13\%).

\section{Validation}

Having shown that our force field is able to correctly reproduce the
\textit{ab-initio} dipoles and forces, we now turn to its validation against
experimental data pertaining to the structure, thermodynamics and
dynamics of aqueous ions at infinite dilution, as well as to the
density of ionic crystals.
We finally investigate the transferability
of the force field to concentrated solutions, which where not taken
into account when ``designing'' the force field.  It is worth pointing
out here that we use experimental data only for validation purposes,
in contrast with all other force fields for aqueous ions, which use
some experimental data for calibration of the parameters.  Out of the
241 parameters defining the force field for the
present set of ions, only 3 (the dispersion damping parameters
$b_D^\textrm{Cl-Cl}$, $b_D^\textrm{Cs-Cl}$ and $b_D^\textrm{Sr-Cl}$)
are determined with the use of experimental data, namely the densities
of the 8 crystals.  In particular, no experimental data on aqueous
ions is used during the calibration process.

\subsection{Simulation details}

For ions at infinite dilution, the system contains a single ion and
215 water molecules in a cubic box of size $L=18.65$~\AA.  For the
crystals, the systems  consist of 256 LiCl,
NaCl, KCl or RbCl, 342 CsCl, 192 MgCl$_2$ or CaCl$_2$, or 256
SrCl$_2$. Systems for concentrated solutions are composed of 27
NaCl, KCl and 458 water molecules in cubic box of sizes 
24.4167~\AA\ and 24.638~\AA, respectively. Electrostatic interactions are computed
using a dipolar Ewald sum~\cite{aguado2003a,laino2008a},
with a tolerance of $1.10^{-7}$ to obtain the self-consistent
dipole moments.  Molecular dynamics in the canonical ensemble are
performed using a Nose-Hoover thermostat with a time constant of
1~ps. 
The system is first equilibrated for 250~ps, and the properties are
determined from subsequent 2.75~ns runs.  The density of the crystals
is determined from simulations in the NPT ensemble at
$P=1$~bar. The thermostat is the same than the one used for the NVT ensemble 
and the barostat is an extension of the one by Martyna \textit{et al.}~\cite{martyna1994a}. 
All simulations are performed using the CP2K simulation package~\cite{cp2k}.

\subsection{Solvation of ions: structure}

\begin{figure}[ht]
\begin{center}
  \includegraphics[scale=0.25]{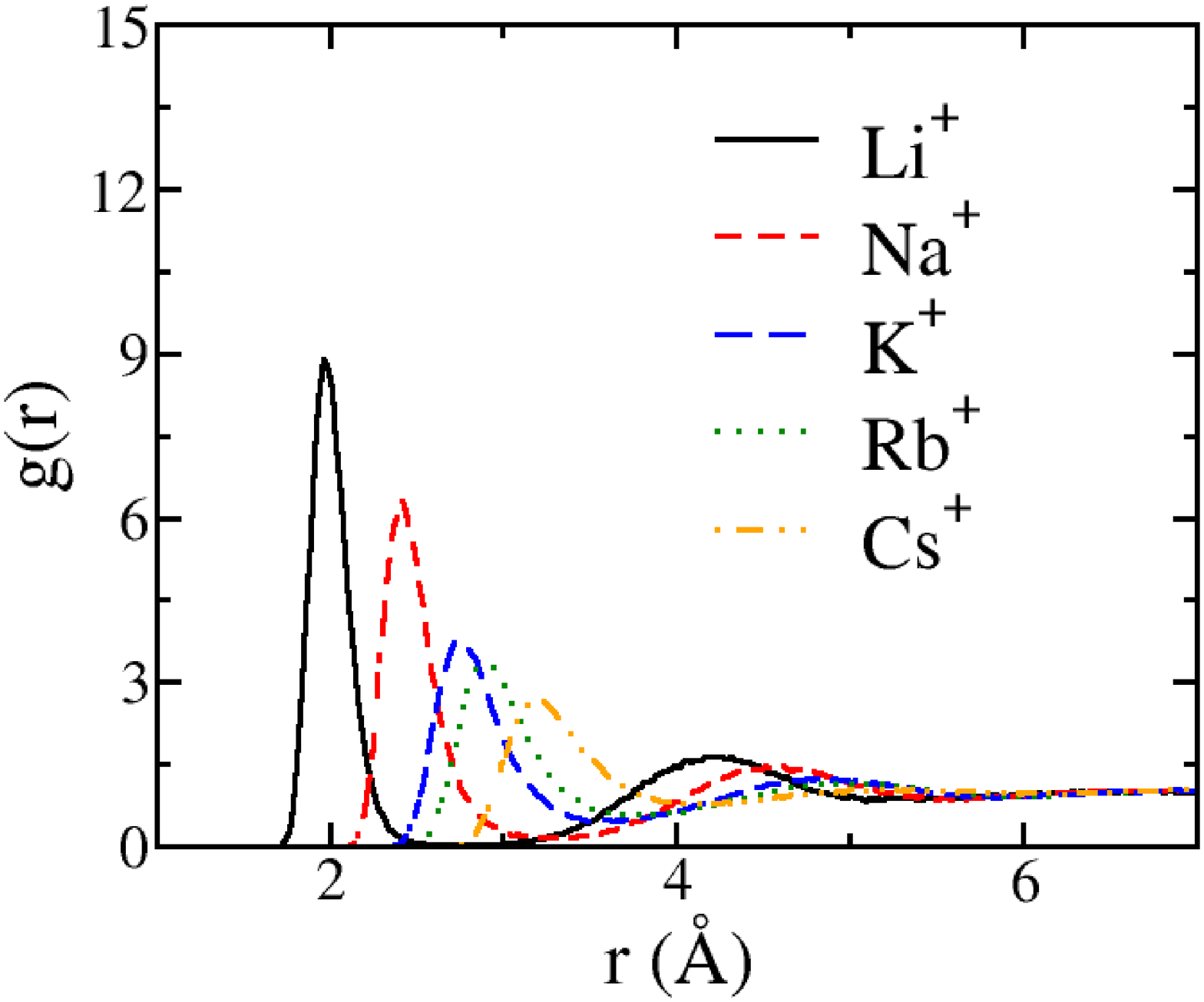}
  \includegraphics[scale=0.25]{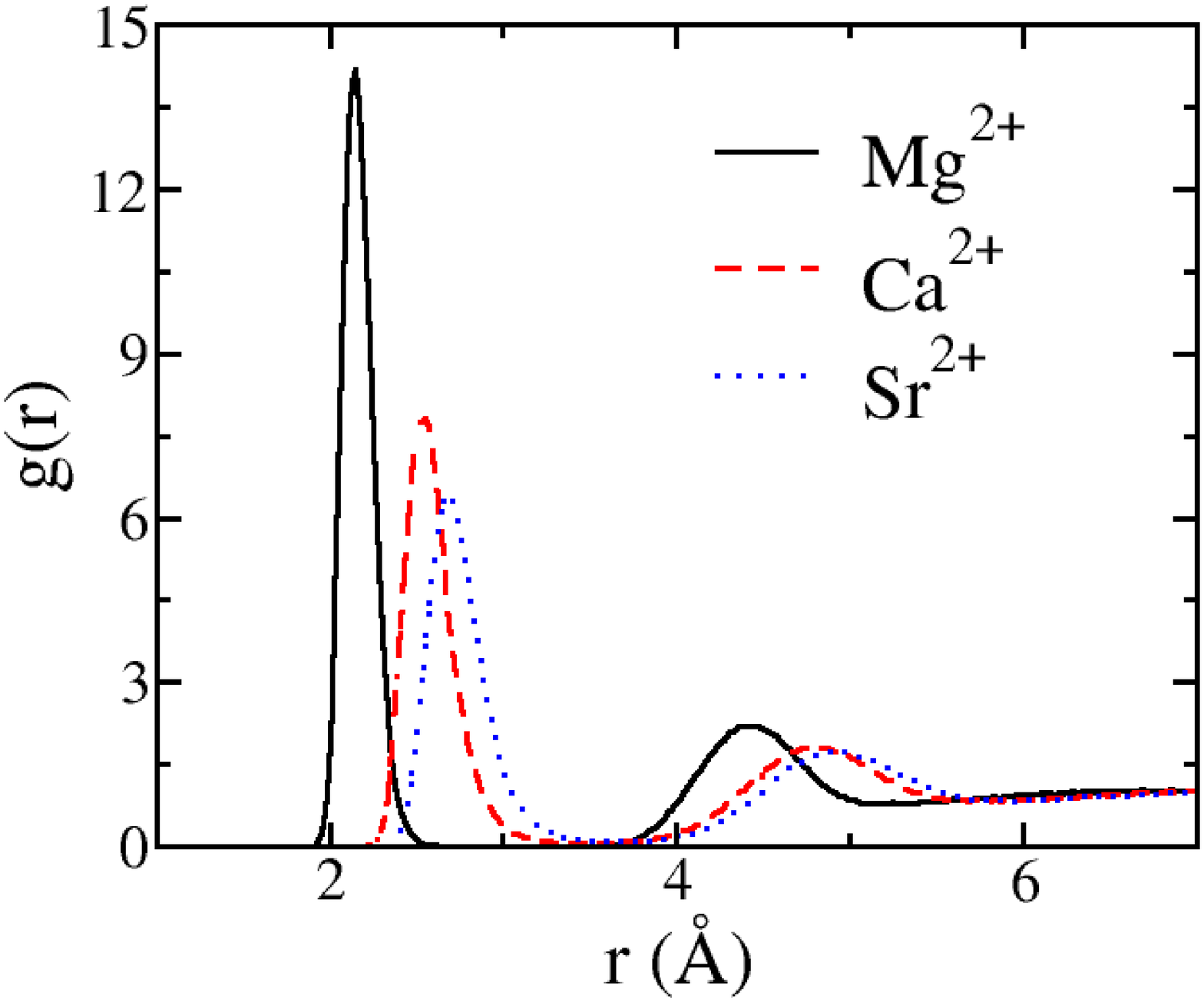}
  \caption{\label{fig:gofr} Ion-oxygen radial distribution functions
    for the aqueous cations.}
\end{center}
\end{figure}

We first inverstigate the structure of the solvation shells around
ions by computing radial distribution functions, reported for
the cations in Fig.~\ref{fig:gofr}. As usually observed, the position
of the first maximum gradually shifts towards larger distances when 
switching from Li$^+$ to Cs$^+$ and from Mg$^{2+}$ to Sr$^{2+}$, while the value
of the maximum decreases and the peak broadens. On the contrary,
moving right along the rows of the periodic table results in a closer
and sharper peak. This arises from the stronger electrostatic
interaction, since the ion-water repulsion remains comparable, as
discussed previously, and reflects a tighter first solvation shell.

The positions of the first maximum and the coordination numbers, defined as
the integral of the ion-O(water) radial distribution function from the origin
out to the first minimum, are
summarized in Table~\ref{tab:static-ppte}, together with the
corresponding experimental values. The value and error estimates of
the coordination numbers are determined from the plateau of the
running values.  Remarkably, all simulated data fall in the reported
experimental ranges.  Particularly encouraging is the agreement with
experimental data for the three divalent ions. While several force
fields are able to correctly predict the position and number of
neighbours for the Mg$^{2+}$ ion, many of them fail to correctly
reproduce that of Ca$^{2+}$. As an example, a force field by Yu 
\textit{et al.} based on a Drude
model of polarizability, which accurately describes the hydration free
energies, predicts a coordination number of 6 for this 
ion~\cite{yu2010a}. Our result is very close to the value of 7.3
obtained with the popular AMOEBA force field used for biomolecular
simulations~\cite{jiao2006a}. The previously available model for
Ca$^{2+}$, with the present water model, predicts a distance of
2.45~\AA, within the experimental range, but it used the EXAFS data of
2.43~\AA\ in the parametrization process. In the case of Sr$^{2+}$, we
find a distance very close to the anomalous X-ray diffraction value of
2.67\AA~\cite{ramos2003a} and a coordination number which is within
the reported experimental range.  

When comparing simulation results for an ion at infinite dilution
with experiments, one should pay attention to the experimental conditions, in
particular the concentration and the nature of the counterion.
For example, Smirnov and Trostin reported an increase of
the Cs$^+$ coordination number with decreasing concentration~\cite{smirnov2007a}.
It is thus not surprising to find our result on the larger side of the
experimental range. Moreover, results using ClO$_4^-$ as a counterion
instead of Cl$^-$ are less likely to be polluted by the formation of ion pairs.
The distances of 2.12 and 2.65~\AA\ between the cation and the nearest water
oxygen, reported with ClO$_4^-$ in Ref.~\cite{ohtaki1993a} for Mg$^{2+}$ and Sr$^{2+}$,
respectively, are in very good agreement with ours (2.13 and 2.68). 

Since most force fields include some experimental data on the structure during
the calibration process~\cite{yu2010a,dang2006a,dang2003a}, it is possible to
obtain a good agreement.  When such data is not included
as a target property, the predicted structure may
not be very accurate. As an example, Horinek \textit{et al.} parametrized a
simple non-polarizable force field optimized for the
simulation of solvation thermodynamics~\cite{horinek2009a}.
The structural properties, used only for validation
purposes, revealed a tendency to underestimate the distances to the
nearest water molecules for cations.
Our results for the chloride ion are very good, as they fall
exactly on the EXAFS value of 2.11~\AA\ determined by Dang \textit{et al.}, whereas
many force fields tend to predict too large a distance for the first
peak~\cite{dang2006a,horinek2009a}, even though they include such
structural properties in the fitting procedure~\cite{dang2006a}.

\begin{table}[ht]
\begin{center}
\begin{tabular}{|l|c|c|c|c|}
  \hline
  Ion & \multicolumn{2}{c|}{Position (\AA)} & \multicolumn{2}{c|}{Coordination Number} \\ \hline 
  &  Sim & Exp       &  Sim        & Exp \\\hline\hline
  Li$^{+}$  & 1.96 & 1.90-2.25 & 4.0         & 4   \\\hline
  Na$^{+}$  & 2.41 & 2.41-2.50 & 5.7$\pm$0.1   & 4-8 \\\hline
  K$^{+}$   & 2.74 & 2.60-2.92 & 6.45$\pm$0.25 & 4-8 \\\hline
  Rb$^{+}$  & 2.88 & 2.80-3.05 & 7.05$\pm$0.25 & 6-8 \\\hline
  Cs$^{+}$  & 3.20 & 2.95-3.21 & 8.3$\pm$0.8   & 6-8 \\\hline\hline
  Mg$^{2+}$ & 2.13 & 2.00-2.15 & 6.0           & 6   \\\hline
  Ca$^{2+}$ & 2.53 & 2.40-2.58 & 7.24$\pm$0.02 & 7-9 \\\hline
  Sr$^{2+}$ & 2.68 & 2.57-2.67 & 7.81$\pm$0.05 & 7.3-10.3\\\hline\hline
  Cl$^{-}$  & 3.11 & 3.05-3.18 & 6.12$\pm$0.12 & 5.3-6.4\\\hline
\end{tabular}
\caption{\label{tab:static-ppte} Structural properties: position of the first maximum 
in radial distribution function and coordination number. The experimental 
values are taken from Refs.~\cite{ohtaki1993a,smirnov2007a,marcus2009a,dang2006a,ramos2003a} }
\end{center}
\end{table}

\begin{figure}[ht]
\begin{center}
\includegraphics[scale=0.35]{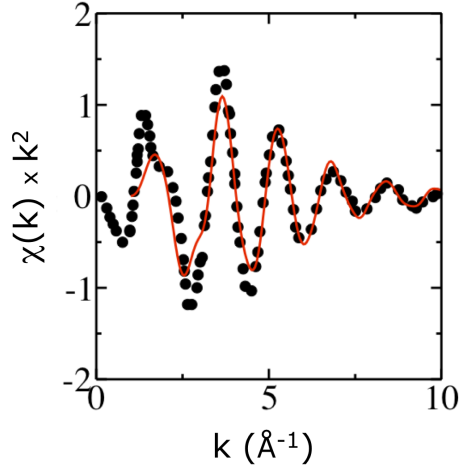}
\caption{\label{fig:caexafs} Comparison between simulated and
  experimental EXAFS~\cite{dang2006a} signal for aqueous Ca$^{2+}$. The agreement is
  very good both for the amplitude, which reflects the number of
  neighbours, and frequency of the oscillations, related to their
  position.  }
\end{center}
\end{figure}

Positions and coordination numbers cannot be measured directly, and
the experimental values are the outcome of a complex numerical
analysis of the raw data, which typically involves several Fourier
transforms and filters which can influence the final result. A more
stringent test of the force field thus consists in comparing the
experimental signal to that obtained by computing the experimental
observables on configurations generated by molecular simulation.  An
example of such a test is given in Fig.~\ref{fig:caexafs}, which
compares the experimental EXAFS signal for aqueous Ca$^{2+}$, obtained
from Ref.~\cite{dang2006a}, to that predicted from our configurations
using the FEFF8 code, which uses an updated version of the Rehr \textit{et
al.} algorithm~\cite{rehr1992a} to evaluate multiple electron
scattering series. The agreement is seen to be very good in the
$k>3$~\AA$^{-1}$ part of the spectrum, both in terms of the amplitude
(which reflect the number of neighbours) and the frequency of
oscillations (related to their position).

\subsection{Solvation of ions: hydration free energy}

Among all the ionic properties one aims to predict, the hydration free
energy $\Delta G_{\text{hyd}}$ is probably the most important, since
it relates to the ability of ions to accomodate their solvation shell
when approaching an interface or other ions. This quantity is almost
always one of the target properties used to design force
fields. Whereas absolute values are difficult to determine,
differences in hydration free energies can be easily computed using a
thermodynamic integration procedure without worrying about the
numerous corrections~\cite{kastenholz2006a,kastenholz2006b} (for
system size, boundary conditions, and the treatment of electrostatic
interactions) needed for the former. The difference $\Delta\Delta
G_\textrm{hyd}\approx\Delta\Delta F_\textrm{hyd}= \Delta
F_\textrm{hyd}^\textrm{K}-\Delta F_\textrm{hyd}^\textrm{Na}$ can be
determined from a thermodynamic path (transmutation) connecting the
systems, by introducing a mixed Hamiltonian $H(\lambda)=\lambda
H_\textrm{K} + (1-\lambda)H_\textrm{Na}$ for $\lambda\in[0,1]$, as
\begin{equation}
\Delta\Delta F = \int_0^1 \left<\frac{\partial H}{\partial
    \lambda}\right>d\lambda  
\end{equation}
For the monovalent ions, we use a $6$-point Gaussian
quadrature\cite{numericalrecipes,intronumericalanalysis} to compute the integral, except for the
Li$^+$-Cs$^+$ transmutation, for which we use an $8$-point
quadrature. Details on this standard quadrature procedure can be found 
in Ref.\cite{numericalrecipes}. In the case of the divalent ions,
where $\partial_\lambda H(\lambda)$ shows a linear variation in
$\lambda$, a simpler trapezoidal rule can be used to approximate the
integral. In this case we used ten equally spaced points (0.1) for
$\lambda_i$ within the interval $[0,1]$.


\begin{table}[ht]
\begin{center}
\begin{tabular}{|l|c|c|c|}
  \hline
  Transmutation & $\Delta\Delta G^\textrm{sim}_\textrm{hyd}$ (kcal/mol) &
  $\Delta\Delta G^\textrm{exp}_\textrm{hyd}$ (kcal/mol) \\ \hline \hline
  Li$^{+}$$\rightarrow$Na$^{+}$ & 26.5 & [23.8;26.2] \\
  Na$^{+}$$\rightarrow$K$^{+}$ & 13.7 & [16.7;17.7] \\
  K$^{+}$$\rightarrow$Rb$^{+}$ & 3.2 & [4.9;5.4] \\
  Rb$^{+}$$\rightarrow$Cs$^{+}$ & 7.6 & [5.5;7.7] \\\hline
  Li$^{+}$$\rightarrow$Cs$^{+}$ & 51.4 & [50.9;57.0] \\\hline\hline
  Mg$^{2+}$$\rightarrow$Ca$^{2+}$ & 82.2 & [77.7;80.3] \\
  Ca$^{2+}$$\rightarrow$Sr$^{2+}$ & 25.3 & [29.8;32.9] \\\hline
  Mg$^{2+}$$\rightarrow$Sr$^{2+}$ & 107.8 & [107.5;113.2] \\\hline
\end{tabular}
\caption{\label{tab:hydfreegibbs} Differences in Gibbs free energy of
  hydration: Simulated and experimental values. The experimental
  values for the monovalent ions are taken from references
  \cite{tissandier1998a,randles1956a,schmid2000a,noyes1962a,gomer1977a,marcus1991a}
  and those for divalent ions from
  \cite{schmid2000a,gomer1977a,marcus1991a}.  }
\end{center}
\end{table}

The hydration free energy 
differences, for all the transmutations considered, are
summarized in Tab.~\ref{tab:hydfreegibbs}, together with the
corresponding experimental values.  The overall agreement with
experiment is very good, with deviations  never
exceeding a few kcal/mol, and the large variations of $\Delta\Delta
G_\textrm{hyd}$ across the ion series being well reproduced.
We note that some force fields are able to reproduce this
quantity slightly more accurately, such as the non-polarizable one of Horinek
\textit{et al.}~\cite{horinek2008a} or the polarizable model (Drude
oscillators) of Yu \textit{et al.}~\cite{yu2010a}.  In these cases, however,
experimental hydration free energies (or differences) 
were used as a target property to calibrate the force field,
whereas we use it here as an independent validation of our \textit{ab-initio}
derived model.

\subsection{Diffusion coefficient}

The diffusion coefficient are computed using from the mean-squared
displacement, as determined by the Einstein relation :
\begin{equation}
D_\textrm{PBC}=\lim\limits_{t \to \infty} \frac{1}{6} 
\frac{ {\rm d} \left< |{\bf r}(t)-{\bf r}(0)|^2\right>}{{\rm d}  t}
\end{equation}
The ``PBC'' subscript emphasizes the fact that the use of periodic
boundary conditions induces a box length dependence on the measured
diffusion coefficient, which takes the form~\cite{yeh2004a}:
\begin{equation}
D_\textrm{PBC} = D_0 - \frac{2.837 k_BT}{6\pi \eta L}
\end{equation}
where $\eta$ is the shear viscosity of the solvent.  For the box
length of $L=18.65$~\AA\ used in our simulations, the correction to
the Dang-Chang water model is approximately
$0.43~10^{-9}$~m$^2$s$^{-1}$ and must not be
neglected ($D_0^{\textrm{H}_2\textrm{O}}=2.72\pm0.09$~m$^2$s$^{-1}$)~\cite{saminpreparation}. For a meaningful comparison with
experiments, we thus extrapolate to the infinite box length limit,
both for the ion and the water diffusion coefficients and compare
the ratios $D_0^\textrm{ion}/D_0^{\textrm{H}_2\textrm{O}}$.

\begin{table}[ht]
\begin{center}
\begin{tabular}{|l|c|c|c|}
  \hline
  Ion       & $(D_0^\textrm{ion}/D_0^{\textrm{H}_2\textrm{O}})^\textrm{sim}$ 
  &$(D^\textrm{ion}/D^{\textrm{H}_2\textrm{O}})^\textrm{exp}$  \\ 
  \hline
  \hline
  Li$^+$    & 0.49 & 0.44 \\
  Na$^+$    & 0.54 & 0.58 \\
  K$^+$     & 0.78 & 0.88 \\
  Rb$^+$    & 0.88 & 0.90 \\
  Cs$^+$    & 0.82 & 0.89 \\
  \hline
  \hline
  Mg$^{2+}$ & 0.31 & 0.31 \\
  Ca$^{2+}$ & 0.35 & 0.34 \\
  Sr$^{2+}$ & 0.35 & 0.34 \\
  \hline
  \hline
  Cl$^{-}$ & 0.71 & 0.88 \\\hline
\end{tabular}
\caption{\label{tab:diff} Ratio between the ion and water diffusion
  coefficients. The experimental values for the ions are taken
  from~\cite{crchandbook2005}, the one for the water
  from~\cite{krynicki1978a}.}
\end{center}
\end{table}

The simulation results are compared to the experimental ratios in
Tab.~\ref{tab:diff}. The relative error is of only 2 to 11\% for the
monovalent cations. The agreement is particularly good for the
divalent cations, for which the relative error does not exceed
3\%. The largest relative error is for the Cl$^-$ ion (19\%).
We performed similar simulations with the force field of 
Dang and co-workers~\cite{dang1992a,chang1997a,dang2002a}, which
uses the same water model. The errors in that case reach 
31\% for Cl$^-$ and 9\% for Ca$^{2+}$ and Sr$^{2+}$.  
Moreover, while our results capture the equal diffusion coefficients
of these two cations, the model of Dang and co-workers underestimates
that of Ca$^{2+}$ and overestimates that of Sr$^{2+}$.

Since force fields generally do not include experimental data on the
dynamics for their calibration, their accuracy for dynamical
properties is usually not as high as for structural and thermodynamic
ones.  The present strategy, which aims at reproducing the forces on
the atoms and molecules, as best as possible, allows us to also
predict the dynamic properties.


\subsection{Crystal density}

\begin{figure}[ht]
\begin{center}
\includegraphics[scale=0.40]{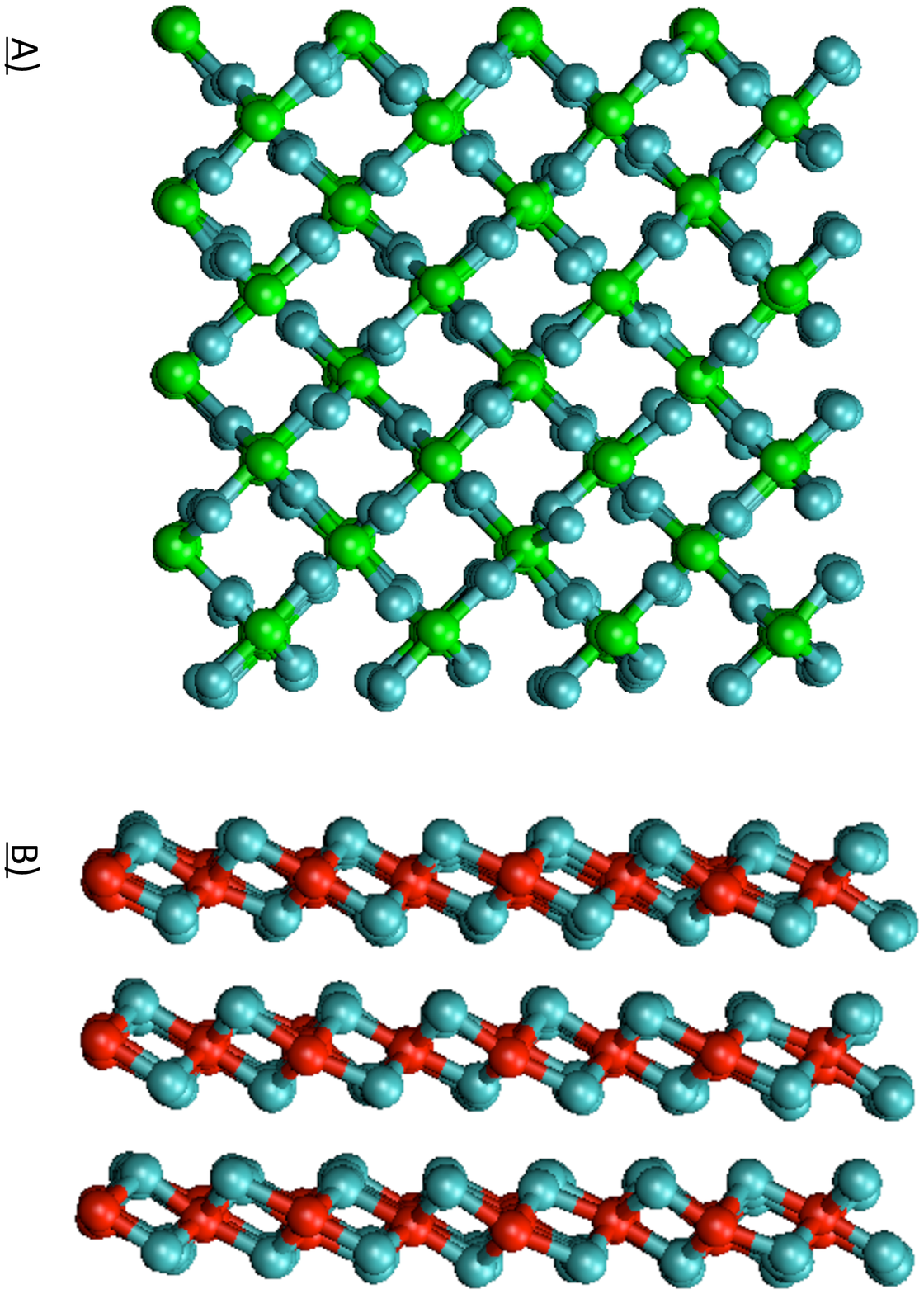}
\caption{\label{fig:cristmgetcl} A) Snapshot of CaCl$_2$ cristal. 
B) Snapshot of MgCl$_2$ cristal. Both structures are stable during the
simulations. 
Cl$^-$ are in cyan, Mg$^{2+}$ in red and Ca$^{2+}$ in green. }
\end{center}
\end{figure}

As a test of the interactions between ions, we now turn to the study
of the ionic crystals. As explained above, out of the 241
parameters defining the force field for the entire family of ions we
have studied, only 3 (the dispersion damping parameters
$b_D^\textrm{Cl-Cl}$, $b_D^\textrm{Cs-Cl}$ and $b_D^\textrm{Sr-Cl}$)
were determined using the experimental densities of the 8 crystals
as target properties. All the systems we studied preserved their 
correct crystal structure during the entire
length of the simulations, even the more complex ones corresponding to
the divalent cations. Fig.~\ref{fig:cristmgetcl} illustrates the
deformed rutile structure of CaCl$_2$ and the lamellar one of MgCl$_2$.
While a complete study of the relative
stability of the different possible phases exceeds the scope of the
present work, this suggests that these phases are at
least metastable.  The simulated densities are compared to the
experimental ones in Tab.~\ref{tab:density}. The overall agreement is
once again good, with relative errors below 10\% except for NaCl
(16\%).

\begin{table}[ht]
\begin{center}
\begin{tabular}{|l|c|c|c|}
  \hline
  Crystal  & $\rho_\textrm{sim}$  (g.cm$^{-3}$) & $\rho_\textrm{exp}$ (g.cm$^{-3}$) \\ 
  \hline                                        
  \hline                                        
  LiCl     & 2.01                               & 2.07                              \\
  NaCl     & 1.83                               & 2.17                              \\
  KCl      & 1.93                               & 1.99                              \\
  RbCl     & 2.98                               & 2.76                              \\
  \hline  \hline                                       
  CsCl     & 4.42                               & 3.99                              \\
  \hline  \hline                                        
  MgCl$_2$ & 2.21                               & 2.33                              \\
  \hline  \hline                                       
  CaCl$_2$ & 2.04                               & 2.15                              \\
  SrCl$_2$ & 3.25                               & 3.05                              \\\hline
\end{tabular}
\caption{\label{tab:density} Density of the crystals at 1~bar and 300~K. 
The experimental values are taken from~\cite{crchandbook2005}.
Note that the correct crystal structures (separated in the table) are preserved
during the simulations. 
}
\end{center}
\end{table}

Transferability to crystals is rarely tested, making comparisons with
other potentials rather difficult. We have again used the force fields
of Dang and coworkers~\cite{dang1992a,chang1997a,dang2002a} to assess
the reliability of our potentials.  Although their potentials give
good results for NaCl (d=2.1~g.cm$^{-3}$), KCl (1.9~g.cm$^{-3}$) and
CsCl (3.8~g.cm$^{-3}$) crystals, the structure proves to be unstable
for LiCl, CaCl$_2$ and SrCl$_2$.  This example shows the need for more
complicated force fields (with more parameters), as they can
provide better transferability.

\subsection{Concentrated solutions}

\begin{figure}[ht]
\begin{center}
\includegraphics[scale=0.3]{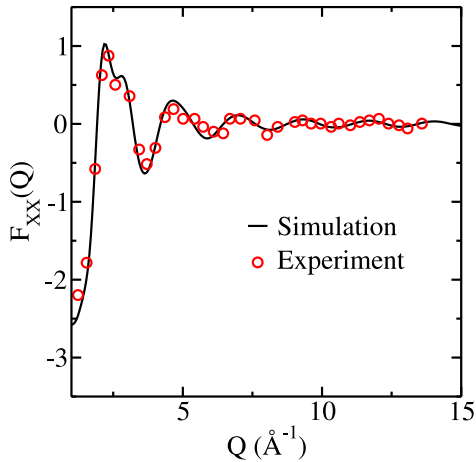}
\caption{\label{fig:nafxx} Comparison between simulated and experimental $F_{XX}(Q)$ 
from Ref.~\cite{mancinelli2007b} for a concentrated NaCl solution 
(one NaCl pair for 17 water molecules).
}
\end{center}
\end{figure}

\begin{figure}[ht]
\begin{center}
\includegraphics[scale=0.3]{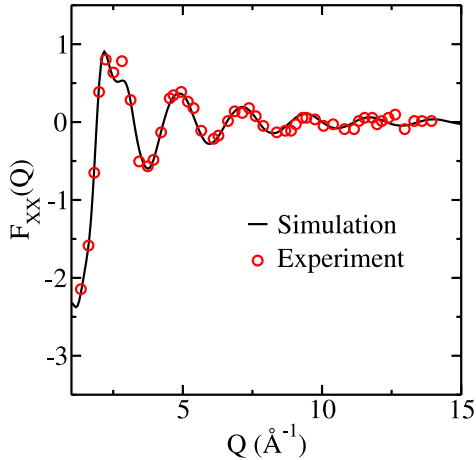}
\caption{\label{fig:kfxx} Comparison between simulated and experimental $F_{XX}(Q)$
from Ref.~\cite{mancinelli2007b} for a concentrated KCl solution 
(one NaCl pair for 17 water molecules).
}
\end{center}
\end{figure}

The previous sections demonstrate the accuracy of the present force
field for both infinitely dilute solutions and crystals. We now test
its transferability to conditions which were not considered during the
construction of the force field, by investigating concentrated ionic
solutions.  We compute the neutron diffraction spectra for
concentrated NaCl and KCl solutions with one ion pair for 17 water molecules
(1:17) from the site-site partial structure factor 
between site $\alpha$ and $\beta$:
\begin{equation}
S_{\alpha\beta}(Q)=4\pi\rho\int r^2(g_{\alpha\beta}(r)-1)\frac{\sin(Qr)}{Qr}dr
\label{eq:sdek1}
\end{equation}
where $\rho$ is the atomic number density of the solution and
$g_{\alpha\beta}(r)$ the corresponding site-site radial distribution
function.  Experimental neutron diffraction allows for the extraction
of composite partial structure factors. We compare our simulations
results to the traditionally used $F_{XX}$ function, defined as:
\begin{equation}
F_{XX}(Q) = \frac{ 
\sum_{\alpha,\beta} (2-\delta_{\alpha\beta}) c_\alpha c_\beta b_\alpha b_\beta S_{\alpha\beta}(Q) 
}{
\left( \sum_{\alpha} c_\alpha b_\alpha \right)^2
}
\label{eq:sdek2}
\end{equation}
where the sums over $\alpha$ and $\beta$ run over all atom types except
hydrogen and $c_\alpha$ and $b_\alpha$ are the atomic fraction and neutron 
scattering length of atom $\alpha$, respectively.
The comparison with the experimental results taken from
Ref.~\cite{mancinelli2007b} in Figs.~\ref{fig:nafxx} and~\ref{fig:kfxx} 
indicates a very good agreement, which confirms the transferability
to concentrated solutions.


\section{Conclusion}

We have shown a successful parametrization of a polarizable force
field for aqueous solutions of Li$^+$, Na$^+$,
  K$^+$, Rb$^+$, Cs$^+$, Mg$^{2+}$, Ca$^{2+}$, Sr$^{2+}$ and Cl$^-$ ions. 
We used the polarizable Dang-Chang model for water and derived all the parameters 
involving ions in the framework of the polarizable ion model of Madden and
co-workers. The procedure relies only on \textit{ab-initio} DFT calculations; 
part of the parameters (polarizabilities, dispersion coefficients) are directly
calculated while the others are extracted from a generalized force- and
dipole-matching procedure. Experimental information is used for validation
purposes only: The structural (first-neighbour distances, coordination numbers),
thermodynamic (hydration free energy differences) and dynamic (diffusion
coefficients) are very well reproduced. The interactions between cations and the
chloride anion are parametrized on calculations performed in the crystal
phases, thus ensuring the accuracy of the force field across the
whole concentration range.

The account of multi-body effects via the polarizability should ensure
a good transferability to more complex conditions: mixtures of these salts,
high temperature and pression~\cite{klotz2009a} and to liquid-vapor 
or liquid-solid interfaces. The next step will consist in extending the
present approach to the interaction of water with the surface of oxide materials.
It will then be possible to use this force field for the study of important problems of
environmental science such as the retention of radionuclides onto clay minerals,
or the water uptake by clays and zeolites.

\acknowledgements

The authors acknowledge financial support from the Agence Nationale de
la Recherche under grant ANR-09-SYSC-012 and from the Groupement
National de Recherches PARIS.
We also would like to thank Pr. Fabio Bruni and Pr. Alan K. Soper for providing us with the raw data
of Ref.~\cite{mancinelli2007b}.


\end{document}